\documentclass[12pt]{iopart}
\usepackage{iopams}
\usepackage{setstack}
\usepackage[dvips]{epsfig,graphics,graphicx}
\usepackage{graphicx}
\usepackage{accents}
\eqnobysec
\def\m@th{\mathsurround=0pt}
\mathchardef\bracell="0365
\def\upbrall{$\m@th\bracell$}
\def\undertilde#1{\mathop{\vtop{\ialign{##\crcr
??? $\hfil\displaystyle{#1}\hfil$\crcr
???? \noalign
???? {\kern1.5pt\nointerlineskip}
???? \upbrall\crcr\noalign{\kern1pt
?? }}}}\limits}

\mathchardef\braceup"0371
\def\upbroll{$\m@th\braceup$}
\def\underhat#1{\mathop{\vtop{\ialign{##\crcr
??? $\hfil\displaystyle{\widehat{#1}}\hfil$\crcr
??? $\hfil\displaystyle{#1}\hfil$\crcr
???? \noalign
???? {\kern1.5pt\nointerlineskip}
???? \upbrall\crcr\noalign{\kern1pt
?? }}}}\limits}
\newcommand{\be}{\begin{equation}}
\newcommand{\ee}{\end{equation}}
\newcommand{\bea}{\begin{eqnarray}}
\newcommand{\eea}{\end{eqnarray}}
\begin{document}
\letter{Time-sliced path integrals with stationary states}
\author{Chris M Field$^\dag$ and Frank W Nijhoff$^\ddag$}
\address{Department of Applied Mathematics, University of Leeds, Leeds LS2 9JT, UK}
\eads{$\dag$ cfield@maths.leeds.ac.uk ~~~~ $\ddag$ nijhoff@maths.leeds.ac.uk}

\begin{abstract}

\noindent
The path integral approach to the quantization of one degree-of-freedom
Newtonian particles is considered within the discrete
time-slicing approach, as in Feynman's original development.
In the time-slicing approximation the quantum mechanical evolution
will generally not have any stationary states.
We look for conditions on the potential energy term such that
the quantum mechanical evolution may possess stationary states
without having to perform a continuum limit.
When the stationary states are postulated to be solutions
of a second-order ordinary differential equation (ODE)
eigenvalue problem it is found that the potential is required to
be a solution of a particular first-order
ODE.  Similarly, when the stationary states are postulated to be solutions
of a second-order ordinary difference equation (O$\Delta$E)
eigenvalue problem the potential is required to be
a solution of a particular first-order O$\Delta$E.
The classical limits (which are at times very nontrivial)
are integrable maps.

\end{abstract}

	\section{Introduction}

The path integral approach to quantum mechanics was developed by R P Feynman
in the middle of the
last century \cite{fey:space}, building on an insight of P M Dirac's \cite{dir:lagran}
(see also \cite{dir:analogy}).
As described by Feynman \cite{fey:Nobel}, the path integral approach ``is another, a third way, of describing quantum mechanics,
which looks quite different to that of Schr\"odinger and Heisenberg, but is equivalent to them.''  There have now been
thousands of papers on the application of path integral techniques to problems in quantum mechanics.
This was recently surveyed in the monograph \cite{grosche:hand}, where results on models that have been explicitly
solved were classified and arranged in tabular form.
In their practical application, however, path integrals are still usually performed via
Feynman's original time-slicing approach (see, for instance, \cite{kleinert:pi}).

This Letter considers the time-sliced path integral of a non-relativistic particle with one degree-of-freedom, as described by a Lagrangian of the form

\begin{equation} \label{CTlag}
\mathcal{L}_O(\dot{q},q) = \frac{1}{2}\dot{q}^2 - V(q),
\end{equation}
and looks for potentials, $V(q)$, such that the quantum evolution may have stationary states whilst still within the discrete time-sliced approximation.

The results have relevance to various areas of research, including the study of lattice effects
in Monte Carlo simulations, asymptotically long-time investigations of quantum systems, and the
investigation of discrete-time quantum systems in their own right.

In section \ref{differential} the potential term is derived under the assumption that the preserved
quantum mechanical stationary states of the discrete-time evolution
are eigenfunctions of a linear second-order ODE.

In section \ref{difference} the potential term is derived under the assumption that the preserved quantum
mechanical stationary states of the discrete-time evolution
are eigenfunctions of a linear second-order O$\Delta$E.

	\section{Differential equation ansatz}\label{differential}

Feynman's approach gives the one-step
time evolution of a wave function $\psi(q_n,t_n)$
within the time-slicing approximation as

	\begin{equation}\label{dtEvln}
		\psi(q_{n+1},t_{n+1}) = N \int_C \exp \!\left[\frac{i}{\hbar} \,
		\varepsilon \, \mathcal{L}(q_{n+1},q_n) \right]\psi(q_n,t_n) d q_n
	\end{equation}
where $C$ is some interval of $\mathbb{R}$, and $n$ is a discrete time parameter.
To obtain the wave function at later times the integrals are concatenated (i.e., iterated).
In Feynman's original paper \cite{fey:space},
the path integral is built up, and worked with throughout the paper, by employing a finite
time interval $\varepsilon = t_{n+1} - t_n$, which is sent to zero at the end of the calculations.  
In this limit (where a finite time interval requires an infinite number of integrals) we obtain
the path integral, see \cite{fey:space}, \cite{fey:quant}.
This `time-slicing' approach is still the method that is
usually employed in the evaluation of path integrals.
As we are studying phenomena within the time-slicing approximation
we set $\varepsilon = 1$ without loss of generality.
The discrete-time Lagrangian is (see \cite{fey:space})

	\begin{equation}\label{LorigG}
		\mathcal{L}(q_{n+1},q_n) = \frac{1}{2}(q_{n+1}-q_n)^2 - V(q_n)
	\end{equation}
(The discrete-time Lagrangian (\ref{LorigG}) follows from the original
Lagrangian (\ref{CTlag}) by the replacement
$\dot{q} = (q_{n+1} - q_n)/\varepsilon$, see \cite{fey:space}.)
For $C = \mathbb{R}$, unitarity shows that
$|N|^2 = (2 \pi \hbar)^{-1}$
(in \cite{fey:space} $N$ is found by considering the $0$th order term of an expansion in the time-step
parameter).  At this stage we refrain from stipulating the range of integration and the
normalization constant $N$.

A simple gauge transformation allows us to consider, instead of (\ref{LorigG}),
the (gauge equivalent)

	\begin{equation}\label{L}
		\mathcal{L}(q_{n+1},q_n) = - q_n q_{n+1} + q_n^2 - V(q_n)
		=: - q_n q_{n+1} - W(q_n).
	\end{equation}

In this section, we look for forms of the potential, $V(q)$, (or, equivalently, $W(q)$) such that after evolution by
equation (\ref{dtEvln}) the new wave function
is a solution to the same eigenvalue problem of a second-order ODE, with the same eigenvalue
(and so will be for any number of evolutions).  
We write the eigenvalue problem as

\begin{equation} \label{MEigen}
M_q \psi(q) = E \psi(q)
\end{equation}
where $M_q$ is an arbitrary linear second-order ODE (which for convenience is written as a partial differential operator)

\begin{equation} \label{MODE}
M_q = p_0(q) \frac{\partial^2}{\partial q^2} + p_1(q) \frac{\partial}{\partial q}
+ p_2(q).
\end{equation}
The new wave function, evolved by (\ref{dtEvln}), although required to be a solution of the same eigenvalue problem,
is not necessarily the same function.  However, the requirement of unitarity will remove the possibility that the
evolved wave function becomes the identically zero solution.

Over-tildes will now be used to denote time updates, i.e., ${\underaccent{\tilde}{q}} = q_{n-1}$, $q := q_n$, $\tilde{q} := q_{n+1}$, etc.  
It is required that

\begin{equation} \label{MtildeEigen}
M_{\tilde{q}} \widetilde{\psi}(\tilde{q}) = E \widetilde{\psi}(\tilde{q}).
\end{equation}
However, by definition

	\begin{eqnarray}
		E \widetilde{\psi}(\tilde{q}) & =  N \int_C \exp \!\left[\frac{i}{\hbar}
		\mathcal{L}(\tilde{q},q) \right] E \psi(q) d q	
		=  N \int_C \exp \!\left[\frac{i}{\hbar} \,
		\mathcal{L}(\tilde{q},q) \right] \bigg\{M_q \psi(q)\bigg\} d q \nonumber\\
		{} & =  N \int_C \bigg\{\overline{M_q} \exp \!\left[\frac{i}{\hbar} \,
		\mathcal{L}(\tilde{q},q) \right]\bigg\} \psi(q) d q + S,
				\label{EPsiUpdate}
	\end{eqnarray}
where $S$ is the surface term

	\begin{equation}
		S = \left[ p_0(q)\exp \!\left[\frac{i}{\hbar} \,
		\mathcal{L}(\tilde{q},q) \right]
		\left(\frac{i}{\hbar} \tilde{q}\psi(q)
	+ \frac{\partial \psi(q)}{\partial q}     \right) \right]_C,
	\end{equation}
and $\overline{M_q}$ is the adjoint of $M_q$.
For (\ref{EPsiUpdate}) to hold it is required that the integral
in equation (\ref{dtEvln}) exists, and, furthermore,
the integral converges uniformly with respect to $\tilde{q}$
in an appropriate region and the integrand of the integral relation
is an analytic function of $\tilde{q}$ in this region when $q$ is
on the line of integration.  It will be assumed that these analytic
requirements hold.
The surface term vanishing, i.e., $S=0$, seems to be generally in
accord with physical requirements (for instance, if $C = \mathbb{R}$, then
$S=0$ is achieved by requiring that $\psi(q)$ and $\frac{\partial \psi(q)}{\partial q}$
tend to zero as $|q| \to \infty$).  If the surface term vanishes then it is required that

	\begin{equation}
		\overline{M_q}\left(\exp \!\left[\frac{i}{\hbar} \,
		\mathcal{L}(\tilde{q},q) \right]\right) =
		M_{\tilde{q}}\left(\exp \!\left[\frac{i}{\hbar} \,
		\mathcal{L}(\tilde{q},q) \right]\right). \label{PDE}
	\end{equation}
This PDE is a functional equation for both $M_q$ and $\mathcal{L}(\tilde{q},q)$.
It can be solved as follows.

Substituting the form of the Lagrangian (\ref{L}) into (\ref{PDE})
and comparing coefficients of different powers of $q$ reveals
$p_0(\tilde{q})$, $p_1(\tilde{q})$ and $p_2(\tilde{q})$ are
quadratic polynomials.  Hence, set

\begin{displaymath}\fl
-\frac{1}{\hbar^2} p_0(\tilde{q}) =
A_0 \tilde{q}^2 + B_0 \tilde{q} + C_0 \qquad
-\frac{i}{\hbar} p_1(\tilde{q}) =
A_1 \tilde{q}^2 + B_1 \tilde{q} + C_1  \qquad
p_2(\tilde{q}) =
A_2 \tilde{q}^2 + B_2 \tilde{q} + C_2 
\end{displaymath}
and substitute this back into (\ref{PDE})
to see how many of the coefficients
are actually independent.

A comparison of the $\tilde{q}^2$ terms of (\ref{PDE}) shows

\be
A_0 q^2 + A_1 q + A_2 = A_0 q^2 + B_0 q + C_0.
\ee
Comparing coefficients of different powers of $q$ reveals $A_1 = B_0$ and $A_2 = C_0$.

A comparison of the $\tilde{q}$ terms of (\ref{PDE}) shows that

\be
	W'(q) = \frac{D q^2 + E q + F}{A_0 q^2 + B_0 q + C_0}
\ee
where we have introduced

\be
D := \frac{1}{2}(A_1 + B_0)  \qquad E := B_1 - i \hbar 2 A_0 \qquad
 F := \frac{1}{2}(B_2 + C_1) - i \hbar B_0.
\ee

A comparison of the $\tilde{q}^0$ terms of (\ref{PDE}) is then made:
The $q^0$ terms reveal that for (\ref{PDE}) to hold, either $B_2 = C_1$
or $F = 0$.  As $F = 0$ limits the form of the potential term, $B_2 = C_1$ is chosen.
The $q^1$ terms reveal that for (\ref{PDE}) to hold $D = A_1$.
However, $A_1 = B_0$ and therefore $D = B_0$.
The $q^2$ terms are trivial.

Taking the above findings together gives the form of $M_q$.
For aesthetic reasons we make the replacements $\alpha := A_0$, $\beta := B_0$, $\gamma := C_0$,
$\epsilon := E$, $\zeta := F$, and set $C_2 = i \hbar \epsilon/2$ so that $M_q$ is formally Hermitian, to obtain
	\begin{equation}\fl
		M_q = - \hbar^2(\alpha q^2+\beta q+\gamma)\frac{\partial^2}{\partial q^2} +
		i \hbar (\beta q^2 + (\epsilon + 2i \hbar \alpha)q + i \hbar \beta + \zeta)\frac{\partial}{\partial q} + \gamma q^2 + (i \hbar \beta + \zeta)q + \frac{i \hbar \epsilon}{2}.
		\label{MqFullMc}
	\end{equation}
The potential term follows from the expression for $W(q)$, and we have

\be \label{WDiff}
	W'(q) = \frac{\beta q^2 + \epsilon q + \zeta}{\alpha q^2 + \beta q + \gamma}
	\qquad\mbox{where}\qquad
	\alpha, \beta, \gamma, \epsilon, \zeta \in \mathbb{R}.
\ee

The quantum evolution given by equations (\ref{dtEvln}), (\ref{L})
and (\ref{WDiff}) (which is defined up to an unimportant constant of unit magnitude)
sends the eigenfunctions of (\ref{MEigen}), where $M_q$ is given by (\ref{MqFullMc}),
to eigenfunctions of the same eigenvalue problem (\ref{MtildeEigen}) with the
same eigenvalue, if the analytic requirements given between
equations (\ref{EPsiUpdate}) and (\ref{PDE}) hold.
If the singularities of the rational function $W'(q)$, (\ref{WDiff}),
(which are the singularities of the ODE given by (\ref{MqFullMc}))
occur on the real-line then there are also issues to be addressed
in the integration of $W'(q)$ to obtain $W(q)$.  (These are issues
associated with the subset of $\mathbb{R}$ on which the model
is defined, which may have singularities of $W'(q)$ as boundaries.
There are numerous possibilities here, which must be dealt with on a
case by case basis.)

With the parameters in generic position, the finite singularities of
the ODE given by (\ref{MqFullMc}) are regular.
According to the classification scheme of \cite{Ronveaux:heun} they have s-rank$=\!1$, while
the singularity at infinity is irregular with s-rank$=\!2$.
Therefore the ODE given by (\ref{MqFullMc}) has s-rank multisymbol $\{1,1,2 \}$; 
hence it is of confluent Heun class.
As a consequence, the analytical work
required to show that the quantum evolution sends an eigenfunction to the
same eigenfunction will be quite involved in general.
One subcase
has already appeared in the literature;
namely the case where
$\alpha = \beta = \zeta = 0$ and $C = \mathbb{R}$.
This is the discrete-time harmonic oscillator, \cite{ber:quantmap}.
In exact analogy with the continuous-time quantum harmonic
oscillator, the eigenfunctions are of the form
$H_k(a q) e^{-b^2 q^2}$, $a, b \in \mathbb{R}$,
where $H_k$ denotes the $k$th Hermite polynomial.  All analytical requirements are met
in this case, and the path integral evolves eigenfunctions to the same eigenfunctions.
That is, these are quantum mechanical stationary states of the system whilst still
within the time-slicing approximation.

The classical limit of the quantum system
given by equations (\ref{dtEvln}), (\ref{L})
and (\ref{WDiff}) is an integrable map of the plane known
as the McMillan map \cite{McMillan:map} (see also \cite{Gram:dpe}).
The classical limit of the invariant is obtained
using $-i \hbar \frac{\partial}{\partial q} = p$.
In this connection it should be noted that integrable quantum mappings, 
in terms of Heisenberg operators, possessing exact quantum operator 
invariants have been proposed before, e.g, in \cite{NPC}, \cite{QN}.

	\section{Difference equation ansatz}\label{difference}

We now consider the quantum mechanical evolution associated with forms of the
potential, $V(q)$, (or, equivalently, $W(q)$) such that
the wave function
is a solution to an eigenvalue problem of a second-order
O$\Delta$E
both before and after its time-evolution within the time-slicing approximation.
Furthermore, the original and evolved wave functions are associated
with the same eigenvalue.
We write the arbitrary linear second-order O$\Delta$E eigenvalue problem as

\begin{equation} \label{NODE}
N_q \psi(q) = E \psi(q)
\qquad
N_q = p_1(q) \exp \left[ a \frac{\partial}{\partial q} \right] + p_0(q)
+ p_{-1}(q) \exp \left[ - a \frac{\partial}{\partial q} \right].
\end{equation}
Define, for convenience, $w(q) :=e^{-\frac{i}{\hbar} W(q)}$.
It is required that

\begin{equation} \label{NtildeEigen}
N_{\tilde{q}} \widetilde{\psi}(\tilde{q}) = E \widetilde{\psi}(\tilde{q}).
\end{equation}
And, by definition

	\be
		E \widetilde{\psi}(\tilde{q}) = N \int_C \exp \!\left[\frac{i}{\hbar} \,
		\mathcal{L}(\tilde{q},q) \right] E \psi(q) d q	.
			\label{EPsiEvlv}
	\ee
It will be assumed that the integral in (\ref{EPsiEvlv}) exists,
and, furthermore, that it converges uniformly with respect to
$\tilde{q}$ for $\tilde{q} \in \mathbb{R}$.
It is also assumed that the integrand is an analytic function of
$\tilde{q}$ for all $q \in \mathbb{R}$.
In this section the usual physical assumptions $C = \mathbb{R}$ and
$\psi(q) \in L^2(\mathbb{R})$ are made, so

	\begin{eqnarray}
		E \widetilde{\psi}(\tilde{q}) & = & N \int_{-\infty}^{\infty} \exp \!\left[-\frac{i}{\hbar}
		q\tilde{q} \right] w(q) \bigg\{
		p_1(q) \psi(q+ a) + p_0(q)\psi(q) + p_{-1}(q) \psi(q - a)   \bigg\}
		d q \nonumber\\
		{} & = & N \int_{-\infty}^{\infty} \exp \!\left[-\frac{i}{\hbar}
		q\tilde{q} \right]
		\bigg\{ w(q-a) p_1(q-a) \exp \!\left[\frac{i}{\hbar}
		a\tilde{q} \right] + w(q) p_0(q)
			\nonumber\\
		{} & {} & \qquad + w(q+a) p_{-1}(q+a) \exp \!\left[-\frac{i}{\hbar}
		a\tilde{q} \right]
		  \bigg\}
		\psi(q) d q.
	\end{eqnarray}
Hence, for (\ref{NtildeEigen}) to hold, it is required that

\begin{eqnarray}
	w(q-a) p_1(q-a) \exp \!\left[\frac{i}{\hbar}
		a\tilde{q} \right] + w(q) p_0(q)
		+ w(q+a) p_{-1}(q+a) \exp \!\left[-\frac{i}{\hbar}
		a\tilde{q} \right]
	 \nonumber\\
	 \, = w(q) \exp \!\left[-\frac{i}{\hbar}
		a q \right] p_1(\tilde{q})
		+ w(q) p_0(\tilde{q})
		+ w(q) \exp \!\left[\frac{i}{\hbar}
		a q \right] p_{-1}(\tilde{q}).
			\label{D22}
\end{eqnarray}
This is a functional equation that is solved
in a similar manner to the development of the previous section
to obtain
$p_{1}(q)$, $p_{0}(q)$, $p_{-1}(q)$ and $w(q)$.
(In this case using the independence of different
powers of $e^{\frac{i}{\hbar} a q}$, rather than $q$.)

After some tedious but straightforward calculations
the following first-order analytic O$\Delta$E for $W(q)$ is obtained,

	\be
		W(q+a) - W(q)
		= i \hbar \ln \Bigg(
		\frac{A_1 \exp \!\left[\frac{i}{\hbar}
		a q \right] + B_1 + C_1 \exp \!\left[-\frac{i}{\hbar}
		a q \right]}{A_{-1} \exp \!\left[\frac{i}{\hbar}
		a q + \frac{i}{\hbar}
		a^2 \right] + A_0 + A_1 \exp \!\left[-\frac{i}{\hbar}
		a q - \frac{i}{\hbar}
		a^2  \right]}\Bigg).
				\label{DiffWDiff}
	\ee
The difference operator (\ref{NODE}), of which the postulated
quantum mechanical stationary states are eigenfunctions,
is

\begin{eqnarray}\fl
			N_q  =  \left\{ A_1 \exp \!\left[\frac{i}{\hbar}
		a q \right] + B_1 + C_1 \exp \!\left[-\frac{i}{\hbar}
		a q \right] \right\} \! \exp \left[ a \frac{\partial}{\partial q} \right]
		+ 
		A_0 \exp \!\left[\frac{i}{\hbar}
		a q \right] + B_0 + B_1 \exp \!\left[-\frac{i}{\hbar}
		a q \right]
		\nonumber\\
		+ \left\{
		A_{-1} \exp \!\left[\frac{i}{\hbar}
		a q \right] + A_0 + A_1 \exp \!\left[-\frac{i}{\hbar}
		a q \right]
		\right\} \!
		 \exp \left[ - a \frac{\partial}{\partial q} \right].
		 		\label{NODEactual}
\end{eqnarray}

The aim of this investigation is to find potentials of one degree-of-freedom
Newtonian systems such that their quantum mechanical time evolution has
stationary states even within the time-slicing approximation.  Hence we wish
to find explicitly real-valued expressions for $W(q)$ (and, consequently,
$V(q)$).  To this end, the following parameter replacements are made,
$A_{-1} = \frac{1}{4}e^{-\frac{i}{\hbar} \frac{a^2}{2} }(\delta + \beta + 2 i \alpha)$, 
$A_{0} = \frac{1}{2}(\zeta + i \gamma)$,
$A_{1} = \frac{1}{4}e^{\frac{i}{\hbar} \frac{a^2}{2} }(\delta - \beta)$,
$B_{1} = \frac{1}{2}(\zeta - i \gamma)$,
$C_{1} = \frac{1}{4}e^{-\frac{i}{\hbar} \frac{a^2}{2}}(\delta + \beta - 2 i \alpha)$;
and substituting into (\ref{DiffWDiff}) gives

	\be
		W(q+a) - W(q)
		= 2 \hbar \tan^{-1} \Bigg(
		\frac{ \beta \sin\left[ \frac{a}{\hbar} \left( q + \frac{a}{2} 				\right)\right]
		+
		\alpha \cos\left[ \frac{a}{\hbar} \left( q + \frac{a}{2} 				\right)\right] + \gamma
		}{
		\delta \cos\left[ \frac{a}{\hbar} \left( q + \frac{a}{2} 				\right)\right]
		- \alpha \sin\left[ \frac{a}{\hbar} \left( q + \frac{a}{2} 				\right)\right] + \zeta
		}\Bigg).
				\label{DiffWDiffreal}
	\ee
The difference operator $N_q$ is also easily rewritten in terms of the new parameters and trigonometric functions.

The
first-order equation obtained by exponentiating (\ref{DiffWDiff})
is easily solved, in a formal manner, by a product of four
hyperbolic gamma functions \cite{Ruij1},
\cite{Ruij2}.
Ruijsenaars' hyperbolic gamma functions \cite{Ruij1} are
so-called `minimal solutions' that are essentially unique,
and have distinctive asymptotic properties (the $w(q)$,
of course, is only defined up to a factor with period $a$).
Alternatively, in the difference case one may consider 
a scenario where the path integral is set up in terms of 
discrete summations, or Jackson $q$-integrals \cite{jackson}, and where the 
first-order O$\Delta$E, (\ref{DiffWDiff}) or (\ref{DiffWDiffreal}), is solved by an iterative procedure.
In a specific case there may
be grounds for choosing a particular solution approach.
This, however, falls outside of the remit of the current
Letter, which stops at giving the equation
that the potential must obey (i.e., (\ref{DiffWDiff}) or (\ref{DiffWDiffreal})).

The classical limit of this system is found
after first making the reverse Schr\"odinger replacement
$-i \hbar \frac{\partial}{\partial q} = p$ (and
$p = -{\underaccent{\tilde}{q}}$ in this gauge).
However, in a na\"ive classical limit $N_q$ trivializes.
On making the replacement $a = a_S \hbar$, where
$a_S$ is independent of $\hbar$, the system now has a rich classical limit,
namely the (integrable) Suris map \cite{Suris:standard}.
The passage to the classical limit is nontrivial; the equation
that the potential must obey is qualitatively changed by the process:
the first-order difference equations
that define $W(q)$ (either (\ref{DiffWDiff}) or, equivalently, (\ref{DiffWDiffreal})) become first-order
differential equations.
From an alternative perspective to that of this Letter,
it can be remarked that we have an `integrable quantization' of the Suris map, where,
to retain integrability, the potential term has nontrivial quantum, $O(\hbar)$, corrections.
(In other settings it has long been realised that to retain integrability
after quantization the potential term may need quantum deformations, see \cite{Jarmo}.)

It is
satisfying to note that in both cases the functional equations
((\ref{PDE}) and (\ref{D22})) give, in a straightforward manner,
the specific form of the second-order differential
or difference operator, as well as the equation
for the potential term, such that the discrete-time
quantum evolution may possess stationary states.

\end{document}